\documentstyle[prl,aps]{revtex}

\begin{document}
\author{Jian Qi Shen\footnote{Electronic address: jqshen@coer.zju.edu.cn}}
\address{Zhejiang Institute of Modern Physics and Department of Physics,
Zhejiang University {\rm (}Yuquan Campus{\rm )}, \\Hangzhou
310027, P.R. China}
\date{\today }
\title{Motion of a charged particle in the magnetic dipole field\footnote{Since the mathematical procedure in this note is trivial, it will be submitted nowhere else for publication, just uploaded at the e-print archives.}}
\maketitle

\begin{abstract}
This note discusses the motion of a charged particle in the
magnetic dipole field and a modified Barut's lepton mass formula.
It is shown that a charged particle in the magnetic dipole filed
has no bound states, which means that Barut's lepton mass formula
may have no physical basis.
\end{abstract}
\pacs{}

Historically, the subquark theories (including the topics such as
the symmetries and structures in leptons and quarks) captured
particular attention of many
researchers\cite{Pati,Terazawa1,Terazawa2,Pati2,Terazawa3,Harari,Shupe,Yasue,Jiao}.
The occurrence of the leptonic fermion chain (e, $\mu$, $\tau$,
...) is a novel phenomenon for which we have so far no theory to
interpret the origin of the generation of leptons. Regarding this
subject, the fundamental problems are as follows: why does the
generation phenomenon exist? what causes the generation number is
not up to three\cite{Dolgov}? Studying the mass formula for
leptons may provide clue to the physicists on how the fundamental
mechanism involved works in the above-mentioned problems. For this
aim, several authors probed the lepton and quark mass
spectra\cite{Barut1,Barut2,Tennakone,Acharge}.

Of all the above investigations, the most important work that
deserves emphasis here is that of Barut\cite{Barut2}, who proposed
a mass formula for muon on the basis of magnetic self-interaction
of the electron\cite{Barut1}. He believed that the radiative
effects give an anomalous magnetic moment to the electron which
implies an extra magnetic energy\cite{Barut1,Barut2}. By using the
quantization formulation according to Bohr-Sommerfeld procedure,
Barut obtained the magnetic energy of a system consisting of a
charge and a magnetic moment as $E_{n}=\lambda n^{4}$ with $n$
being a principal quantum number. Here the coefficient
$\lambda=\frac{3}{2\alpha}$ with $\alpha$ denoting the
electromagnetic fine structure constant. The lepton mass formula
suggested by Barut is\cite{Barut2}
\begin{equation}
m_{n}=\left(1+\frac{3}{2\alpha}\sum_{l=0}^{n}l^{4}\right)m_{\rm
e},    \label{eq100}
\end{equation}
where $m_{\rm e}$ stands for the electron mass and the integer $n$
represents the generation label of charged leptons (for instance,
$n=0, 1,2$ corresponds to the electron, muon ($\mu$) and tau
($\tau$) particle, respectively). It should be noted that, as far
as the known fermion chain (e, $\mu$, $\tau$) is concerned, this
mass formula of charged leptons agrees with experimental values
extremely well.

Barut argued that although the Bohr-Sommerfeld quantization is
approximative, the final result might be exact as was the case in
Bohr-Sommerfeld derivation of the Balmer formula\cite{Barut2}.
Here, however, I will show that this viewpoint may be not the true
case. In this note, by considering the motion of a charged
particle in the field of a magnetic dipole, it will be shown that
the charged particle near a magnetic dipole may have no bound
states. This, therefore, implies that Barut's magnetic energy
formula $E_{n}=\lambda n^{4}$ may be not valid.

In history, Dirac investigated the motion of an electron in the
field of a magnetic monopole and showed that the electron cannot
be bound to the monopole field\cite{Dirac}. Even though Dirac did
not take into account the electron spin (which will give rise to a
magnetic moment) in his analysis, the above conclusion is still
valid if one treats the same problem by taking into consideration
the electron spin (and hence its spinning magnetic
moment)\cite{Chandra}. In what follows I will deal with the
problem of the stationary Klein-Gordon equation governing the
motion of a scalar particle (or an electron without taking account
of its spin and spinning magnetic moment) in the presence of a
magnetic dipole field.

It is well known that the three-dimensional magnetic vector
potential of a magnetic dipole field written in a polar coordinate
system ($r, \theta, \varphi$) is $A_{\varphi}=\lambda\frac{r\sin
\theta}{\left(r^{2}+a^{2}\right)^{\frac{3}{2}}}$, where the
coefficient is $\lambda=\frac{\mu_{0}Ia^{2}}{4}$. Here, $\mu_{0}$,
$a$ and $I$ denote the vacuum permeability, ring radius and
electric current of the magnetic dipole. Note that here it is
assumed that a circular ring carrying a current $I$ forms a
magnetic dipole. If the radius $a$ is much less than the spatial
scale, then the above-mentioned magnetic vector potential can be
reduced to $A_{\varphi}=\lambda\frac{\sin \theta}{r^{2}}$. For
convenience, in this note we consider only the 2D case ({\it
i.e.}, $\theta=\frac{\pi}{2}$) of the stationary Klein-Gordon
equation, which is of the form

\begin{equation}
\frac{1}{\rho}\frac{\partial}{\partial
\rho}\left(\rho\frac{\partial \psi}{\partial
\rho}\right)+\frac{1}{\rho^{2}}\left(\frac{\partial}{\partial
\varphi}-\frac{ie}{\hbar}\rho
A_{\varphi}\right)^{2}\psi=\left[-\frac{E^{2}}{\hbar^{2}c^{2}}+\left(\frac{mc}{\hbar}\right)^{2}\right]\psi
\label{eq1}
\end{equation}
in the 2D polar coordinate system ($\rho, \varphi$). Substitution
of $\psi=R(\rho)\exp [im\varphi]$ into (\ref{eq1}) yields
\begin{equation}
\left(\frac{{\rm d}^{2}}{{\rm d}\rho^{2}}+\frac{1}{\rho}\frac{{\rm
d}}{{\rm
d}\rho}-\frac{m^{2}}{\rho^{2}}\right)R+\frac{\eta}{\rho^{3}}R+\frac{\sigma}{\rho^{4}}R+{\mathcal
E}R=0          \label{eq2}
\end{equation}
with ${\mathcal
E}=\frac{E^{2}}{\hbar^{2}c^{2}}-\left(\frac{mc}{\hbar}\right)^{2}$,
$\eta=\frac{2em\lambda}{\hbar}$ and
$\sigma=-\frac{e^{2}\lambda^{2}}{\hbar^{2}}$. Similar to the
discussion of 2D hydrogen atom, we insert
\begin{equation}
R(\rho)=\rho^{|m|}\exp (-\beta \rho)u(\rho)   \quad {\rm with}
\quad  \beta=\sqrt{-{\mathcal E}}
\end{equation}
into Eq.(\ref{eq2}), and obtain
\begin{equation}
\frac{{\rm d}^{2}u}{{\rm
d}\rho^{2}}+\left[2\left(\frac{|m|}{\rho}-\beta\right)\frac{{\rm d
}u}{{\rm d}\rho}+\frac{1}{\rho}\frac{{\rm d }u}{{\rm
d}\rho}\right]+\left[\frac{\eta}{\rho^{3}}-\frac{\beta}{\rho}\left(2|m|+1\right)+\frac{\sigma}{\rho^{4}}\right]u=0.
\label{eq3}
\end{equation}
On substituting $u=\sum^{\infty}_{\nu=0}b_{\nu}\rho^{s+\nu}$ into
Eq.(\ref{eq3}), making use of $\frac{{\rm d}^{2}u}{{\rm
d}\rho^{2}}=\sum^{\infty}_{\nu=0}b_{\nu}(s+\nu)(s+\nu-1)\rho^{s+\nu-2}$,
$\frac{{\rm d}u}{{\rm
d}\rho}=\sum^{\infty}_{\nu=0}b_{\nu}(s+\nu)\rho^{s+\nu-1}$,
$\frac{1}{\rho}\frac{{\rm d}u}{{\rm
d}\rho}=\sum^{\infty}_{\nu=0}b_{\nu}(s+\nu)\rho^{s+\nu-2}$ and
equating the coefficients of the various powers of $r$ to zero we
get the recurrence relation
\begin{equation}
b_{\nu+1}(s+\nu+1)(s+\nu)-2\beta
b_{\nu}(s+\nu)+(2|m|+1)b_{\nu+1}(s+\nu+1)-\beta(2|m|+1)b_{\nu}+\eta
b_{\nu+2}+\sigma b_{\nu+3}=0.           \label{eq4}
\end{equation}

For simplicity, we assume that in Eq.(\ref{eq4}) the $\sigma$-
term is much less than the $\eta$- term, so the former one can be
ignored in what follows. Thus the recurrence relation (\ref{eq4})
without the $\sigma$- term is rewritten as
\begin{equation}
b_{\nu+1}(s+\nu+1)(s+\nu+2|m|+1)=\beta(2s+2\nu+2|m|+1)b_{\nu}-\eta
b_{\nu+2}.            \label{eq5}
\end{equation}
When $\nu$ tends to $+\infty$, $b_{\nu+1}\rightarrow \frac{2\beta
b_{\nu}}{\nu+1}-\frac{\eta b_{\nu+2}}{(\nu+1)^{2}}$. This,
therefore, implies that $b_{\nu+1}\rightarrow \frac{1}{(\nu+1)!}$
and the behavior of $u$ converges like $\exp(2\beta\rho)$ for
$\nu\rightarrow \infty$, which is divergent at $\rho\rightarrow
\infty$. So, the maximal $\nu$ in the series of the function $u$
should be a finite integer rather than an infinite one. In the
following we will discuss the recurrence relation from the various
aspects:
\\ \\

\textbf{A.}

We set $q_{\nu}=(s+\nu+1)(s+\nu+2|m|+1)$,
$p_{\nu}=2s+2\nu+2|m|+1$, and from Eq.(\ref{eq5}) we can obtain
\begin{eqnarray}
q_{\nu}b_{\nu+1}&=&\beta p_{\nu}b_{\nu}-\eta b_{\nu+2},
\nonumber \\
q_{\nu+1}b_{\nu+2}&=&\beta p_{\nu+1}b_{\nu+1}-\eta b_{\nu+3}.
\label{eq66}
\end{eqnarray}
It follows from Eqs.(\ref{eq66}) that
\begin{equation}
\left(q_{\nu}+\eta\frac{p_{\nu+1}}{q_{\nu+1}}\right)b_{\nu+1}=\beta
p_{\nu}b_{\nu}+\frac{\eta^{2}}{q_{\nu+2}}b_{\nu+3}. \label{eq6}
\end{equation}
If it is required that $b_{\nu+1}=0$ (and hence $b_{\nu+n}=0$,
$n>1$), then only the case of $\beta p_{\nu}=0$ satisfies this
requirement, since $b_{\nu+3}=0$. This, therefore, means that
$\beta=0$. Thus ${\mathcal E}=0$. But unfortunately the solution
corresponding to such an ${\mathcal E}$ is not a bound one.

 \textbf{B.}

If we assume that $b_{\nu}=0$, $b_{\nu+3}=0$, then it follows from
Eqs.(\ref{eq66}) that
\begin{eqnarray}
q_{\nu}b_{\nu+1}+\eta b_{\nu+2}=0,                  \nonumber \\
\beta p_{\nu+1}b_{\nu+1}-q_{\nu+1}b_{\nu+2}=0.  \label{eq666}
\end{eqnarray}
If one requires Eqs.(\ref{eq666}) to possess nonvanishing
$b_{\nu+1}$, $b_{\nu+2}$, then he can arrive at
\begin{equation}
q_{\nu}q_{\nu+1}+\eta\beta p_{\nu+1}=0,   \label{eq7}
\end{equation}
namely, the determinant of the coefficient matrix of
Eqs.(\ref{eq666}) is vanishing. This, however, means that
$\beta\leq 0$ and that it will cause the function $u$ to be
divergent at large spatial scale ($r\rightarrow \infty$).

\textbf{C.}

According to Eqs.(\ref{eq66}), we can obtain
\begin{eqnarray}
q_{\nu}b_{\nu+1}&=&\beta p_{\nu}b_{\nu}-\eta b_{\nu+2},                  \nonumber \\
q_{\nu+1}b_{\nu+2}&=&\beta p_{\nu+1}b_{\nu+1}-\eta b_{\nu+3},
                  \nonumber \\
                  q_{\nu+2}b_{\nu+3}&=&\beta p_{\nu+2}b_{\nu+2}-\eta
                  b_{\nu+4}.
\end{eqnarray}
If, for example, $b_{\nu+3}=0$, $p_{\nu+2}=0$, it follows that
$b_{\nu+4}=0$, and consequently $b_{\nu+5}=0$, $b_{\nu+6}=0$. So,
we have
\begin{equation}
q_{\nu}b_{\nu+1}=\beta p_{\nu}b_{\nu}-\eta b_{\nu+2},   \quad
q_{\nu+1}b_{\nu+2}=\beta p_{\nu+1}b_{\nu+1}.   \label{eq9}
\end{equation}
Under the condition $b_{\nu}\neq 0$, it follows that
\begin{equation}
\left(q_{\nu}+\frac{\eta
p_{\nu+1}}{q_{\nu+1}}\right)b_{\nu+1}=\beta p_{\nu}b_{\nu}.
\label{eq10}
\end{equation}
So, we have the following recurrence relations
\begin{equation}
b_{\nu+1}=\frac{\beta p_{\nu}}{q_{\nu}+\frac{\eta
p_{\nu+1}}{q_{\nu+1}}}b_{\nu},  \qquad b_{\nu+2}=\frac{\beta^{2}
p_{\nu+1}p_{\nu}}{q_{\nu}q_{\nu+1}+\eta p_{\nu+1}}b_{\nu}, \quad
...
 \label{eq11}
\end{equation}
It is apparently seen that all the coefficients $b_{\nu+n}$ can be
expressed in terms of $b_{\nu}$ and $\beta$. Unfortunately, it is
not possible for us to obtain $\beta$.
\\ \\

In view of the above discussions, it can be easily seen that the
series $u$ cannot terminate for a positive $\beta$. Thus a charged
particle is never bound to the magnetic dipole filed. This,
therefore, means that Barut's lepton mass formula\cite{Barut2}
lacks a power physical basis.  Even though my result is obtained
only by considering the motion of a charged scalar particle in a
magnetic field of a magnetic doublet, it may be concluded that
this result may also hold for a charged spinning particle ({\it
e.g.}, electron) in the presence of the magnetic dipole field, by
analogy with Dirac and Chandra's work\cite{Dirac,Chandra}.
\\ \\

A great majority of the lepton mass formulae have two
disadvantages: (i) they cannot agree with experimental results
very well; (ii) the generation number of fermions in these mass
formulae are often infinite and/or they cannot interpret the
finite-generation-number phenomenon. Although Barut's formula is
in good agreement with experiments (agreeing to the experimental
results about one part in $10^{3}$), it cannot explain that why
the generation number is finite [This can be seen from
Eq.(\ref{eq100}), where the generation label $n$ can take the
arbitrary integers]. In this note, I will put forward a potential
mass formula for the charged leptons, which may overcome the two
above-mentioned disadvantages. Such a mass formula is written as
follows\footnote{This lepton mass formula was suggested in
Nov.1999 - Mar.2000. In fact, such an expression for the leptonic
mass spectrum may be considered the modified version of Barut's
lepton mass formula (\ref{eq100}).}
\begin{equation}
m_{n}=\left(1+\frac{1}{2\alpha}\sum_{l=0}^{n}C_{3}^{l}l^{4}\right)m_{\rm
e }    \quad    {\rm with}  \quad C_{3}^{l}=\frac{3!}{l!(3-l)!}.
\label{eq12}
\end{equation}
Here $n$ denotes the generation label of various charged leptons.
It should be emphasized that the complete agreement exists between
this lepton mass spectrum (\ref{eq12}) and experiments to an
incredibly high degree of accuracy ({\it i.e.}, agreeing to the
experiments about one part in $10^{3}$).

Another interesting lepton mass formula suggested by us is written
in the following form\cite{Shen}

\begin{equation}
m_{n}=C_{3}^{n}\left(\frac{1}{2}\right)^{n^{2}}\left(\frac{1}{\alpha}\right)^{n}m_{\rm
e}   \quad    {\rm with}  \quad C_{3}^{n}=\frac{3!}{n!(3-n)!},
\label{eq13}
\end{equation}
which also agrees reasonably well with experimental values for the
masses of e, $\mu$ and $\tau$.

Apparently, both the above two lepton mass spectra (\ref{eq12})
and (\ref{eq13}) accommodate four generations of fermions, since
the generation label $n$ can take only 0, 1, 2 and 3, which
correspond to electron, muon, tau and f particle\footnote{The mass
of the charged lepton corresponding to $n=3$, which we call, for
brevity, f lepton, is respectively $3.4$ and $2.6$ Gev in
Eq.(\ref{eq12}) and (\ref{eq13}). Note that here the name of the f
lepton (should such exist) is inspired by considering that the f
lepton may be the {\it fourth} (and even the {\it final})
generation of leptons in accordance with the mass spectra
(\ref{eq12}) and (\ref{eq13}).}.

Additionally, I can also suggest an other mass formula for charged
leptons\footnote{This mass formula was proposed in the spring of
2000. It has not yet been published elsewhere.}
\begin{equation}
m_{n}=C_{3}^{n}C_{2}^{n}\left(\frac{1}{4\alpha}\right)^{n}m_{\rm
e}      \quad    {\rm with}  \quad C_{2}^{n}=\frac{2!}{n!(2-n)!},
\label{eq14}
\end{equation}
which accommodates only three generations of fermions. This
formula agrees with experimental values to less than $1\%$.

All the above mass spectra (\ref{eq12})-(\ref{eq14}) for the
charged leptons are established based on both the experimental
values and the mathematical structures of the so-called
supersymmetric Pegg-Barnett oscillator\cite{Shen}. It is believed
that the investigation of the leptonic mass spectra may provide us
with an insight into the generation problem (generation
phenomenon, generation origin and generation structures) of
leptons and quarks. We hold that these leptonic mass spectra
presented in this note deserve further investigation both
theoretically and experimentally.
\\ \\

\textbf{Acknowledgements}  This work was supported in part by the
National Natural Science Foundation of China under Project No.
$90101024$.

\end{document}